\documentclass[twocolumn,showpacs,preprintnumbers,amsmath,amssymb]{revtex4}

\usepackage{graphicx}
\usepackage{dcolumn}
\usepackage{bm}
\usepackage{psfrag}
\usepackage{amsmath}
\usepackage{subfigure}
\usepackage{float}

\begin{document}

\preprint{APS/123-QED}

\title{Population and phase dynamics of $F=1$ spinor condensates \\ in an external magnetic field}

\author{D. R. Romano}
\author{E. J. V. de Passos}%
\affiliation{%
Instituto de F\'{\i}sica, Universidade de S\~ao Paulo \\
Caixa Postal 66318, CEP 05315-970, S\~ao Paulo, S\~ao Paulo, Brasil 
}%

\date{\today}

\begin{abstract}
We show that the classical dynamics underlying the mean-field description of homogeneous mixtures of spinor $F=1$ Bose-Einstein condensates in an external magnetic field is integrable as a consequence of number conservation and axial symmetry in spin space. The population dynamics depends only on the quadratic term of the Zeeman energy and on the strength of the spin-dependent term of the atom-atom interaction. We determine the equilibrium populations as function of the ratio of these two quantities and the miscibility of the hyperfine components in the ground state spinors are thoroughly discussed. Outside the equilibrium, the populations are always a periodic function of time where the periodic motion can be a libration or a rotation. Our studies also indicates the absence of metastability.
\end{abstract}

\maketitle

\section{\label{sec:level1}Introduction}

The recently realized trapping of sodium atoms by purely optical means \cite{Stamper} opens up the possibility of studying "spinor" condensates in which the spin degrees of freedom are not frozen \cite{Stenger}.

Many authors have investigated, in the framework of Bogoliubov theory, the ground state configurations and the low-lying collective excitations of homogeneous mixtures of $F=1$ spinor condensates in the absence \cite{Ho} and in the presence of an external magnetic field \cite{Stenger,Ohmi}. These studies predicted a variety of new phenomena such as the existence of spin domains in the ground state \cite{Stenger,Isoshima} and the propagation of spin waves \cite{Ho,Ohmi}.

The mean-field dynamics inherent in these works is known to be equivalent to a classical dynamics whose degrees of freedom are the phase and population of the three hyperfine levels \cite{Ho,Ohmi}.

In this paper we show that this classical dynamics is integrable as a consequence of number conservation and axial symmetry in spin space. By a proper  canonical transformation it reduces to a dynamics involving only one degree of freedom. The determination of the equilibrium points reveals a rich structure in phase-space. Contour plots of the constant energy surfaces show that the population is a periodic function of time, where the periodic motion can be a libration or a rotation.

Our studies differs from references \cite{Stenger,Ho,Ohmi} by taking explicitly into account the constraint of the axial symmetry in spin space. This allow us to make a detailed discussion of the properties of the equilibrium configurations and of the population and phase dynamics of the spinor condensate, which complements previous works \cite{Stenger,Ho,Ohmi}.

\section{\label{sec:level2}The equations of motion}

The hamiltonian of our system of $F=1$ homogeneous mixture of bosonic atoms in the presence of an external magnetic field is equal to \cite{Stenger,Ho,Ohmi}

\begin{eqnarray}
&\hat{H}&=\sum_{\alpha,\vec{k}}\epsilon_{\vec{k}}a_{\alpha\vec{k}}^{\dagger}a_{\alpha\vec{k}}-\frac{p}{\hbar}\sum_{\alpha,\vec{k}}\langle\alpha|\hat{S}_{z}|\alpha\rangle a_{\alpha\vec{k}}^{\dagger}a_{\alpha\vec{k}} \nonumber \\
&+&\frac{q}{\hbar^{2}}\sum_{\alpha,\vec{k}}\langle \alpha|\hat{S}_{z}^{2}|\alpha \rangle a_{\alpha\vec{k}}^{\dagger}a_{\alpha\vec{k}} \nonumber \\
&+&\frac{c_{0}}{2V}\sum_{\substack{\vec{k}_{1},\vec{k}_{2}\\ \vec{q},\alpha,\beta}}a_{\alpha\vec{k}_{1} +\vec{q}}^{\dagger}a_{\beta\vec{k}_{2}-\vec{q}}^{\dagger}a_{\beta\vec{k}_{2}}a_{\alpha\vec{k}_{1}}  \\
&+&\frac{c_{2}}{2V}\sum_{\substack{\vec{k}_{1},\vec{k}_{2}\\ \vec{q}, \alpha,\beta \\ \alpha',\beta'}}\langle \alpha|\hat{\vec{S}}|\alpha'\rangle\cdot \langle\beta|\hat{\vec{S}}|\beta'\rangle a_{\alpha\vec{k}_{1}+\vec{q}}^{\dagger}a_{\beta\vec{k}_{2}-\vec{q}}^{\dagger}a_{\beta'\vec{k}_{2}}a_{\alpha'\vec{k}_{1}} \nonumber 
\label{equation1}
\end{eqnarray}

\noindent with:

\begin{eqnarray}
\epsilon_{\vec{k}}=\frac{\hbar^{2}\vec{k}^{2}}{2m} \nonumber
\end{eqnarray}

In (\ref{equation1}), $a_{\alpha\vec{k}}^{\dagger}$ creates an atom in the hyperfine level $\alpha$, $\alpha=1,0,-1$, with momentum $\hbar\vec{k}$, $p$ and $q$ are the intensities of the linear and quadratic terms of the Zeeman energy \cite{Stenger} and $c_{0}$ and $c_{2}$ are, respectively, the strengths of the spin-independent and spin-dependent terms of the atom-atom interaction \cite{Stenger,Ho,Ohmi}.

The hamiltonian (\ref{equation1}) is number conserving and axially symmetric in spin space,

\begin{eqnarray}
\left[ \hat{H},\hat{N}\right]=0, \left[ \hat{H},e^{-i\frac{\phi \hat{S}_{z}}{\hbar}}\right]=0 \nonumber
\label{equation2}
\end{eqnarray}

In the mean-field theory we suppose that the condensate is a coherent combination of atoms in the $\vec{p}=0$ state,

\begin{eqnarray}
|\vec{z}\rangle = e^{-\frac{1}{2}\sum_{\alpha}|z_{\alpha}|^{2}} e^{\sum_{\alpha}z_{\alpha}a_{\alpha 0}^{\dagger}} |0\rangle
\label{equation3}
\end{eqnarray}

\noindent where $|0\rangle$ is the vacuum. The complex numbers $z_{\alpha}$ are the condensate wave-functions for the atoms in the hyperfine level $\alpha$. To find the time evolution of $z_{\alpha}$, we use the time-dependent variational principle \cite{Ho,Dalfovo}

\begin{eqnarray}
\delta S=\delta\int i\hbar\langle\vec{z}|\dot{\vec{z}}\rangle-\langle\vec{z}|\hat{H}|\vec{z}\rangle   dt=0
\label{equation3}
\end{eqnarray}

\noindent which reduces to

\begin{eqnarray}
\delta S=\delta\int \left( \sum_{\alpha}i\hbar z_{\alpha}^{*}\dot{z_{\alpha}}-H_{0}(\vec{z},\vec{z}^{*})\right) dt=0
\label{equation4}
\end{eqnarray}

\noindent where the hamiltonian $H_{0}(\vec{z},\vec{z}^{*})$ is given by

\begin{eqnarray}
H_{0}(\vec{z},\vec{z}^{*})&=&\langle \vec{z}|\hat{H}|\vec{z}\rangle = -p\left( |z_{1}|^{2}-|z_{-1}|^{2}\right) \nonumber \\
&+&q\left( |z_{1}|^{2}+|z_{-1}|^{2}\right) + \frac{c_{0}}{2V}\left( \sum_{\alpha}|z_{\alpha}|^{2}\right)^{2} \nonumber \\
&+&\frac{c_{2}}{2V}\left( (|z_{1}|^{2}-|z_{-1}|^{2})^{2}\right. \nonumber \\
&+&\left. 2|z_{0}|^{2}(|z_{1}|^{2}+|z_{-1}|^{2}) \right. \nonumber \\ 
&+& \left. 2z_{1}^{*}z_{-1}^{*}z_{0}^{2}+2z_{1}z_{-1}(z_{0}^{*})^{2}\right) \nonumber \\
\label{equation5}
\end{eqnarray}

Imposing that the action is stationary with respect to variations of $z_{\alpha}$, we get Hamilton equations of motion in complex coordinates

\begin{eqnarray}
i\hbar \dot{z}_{\alpha}=\frac{\partial H_{0}}{\partial z_{\alpha}^{*}}, \,\,\,\,\,\,\,\,\,  -i\hbar \dot{z}_{\alpha}^{*}=\frac{\partial H_{0}}{\partial z_{\alpha}}
\label{equation6}
\end{eqnarray}

To take advantage of number conservation and axial symmetry in spin space we perform two canonical transformations. The first one introduces, as canonical variables in phase-space, the population and phase of each hyperfine component by the transformation

\begin{eqnarray}
z_{\alpha}=\sqrt{N_{\alpha}}e^{-i\theta_{\alpha}}
\label{equation7}
\end{eqnarray}

\noindent such that the action (\ref{equation4}) reduces to

\begin{eqnarray}
S=\int_{t_{1}}^{t_{2}}\left( \hbar\sum_{\alpha}N_{\alpha}\dot{\theta_{\alpha}}-H_{0}(\vec{N},\vec{\theta})\right) dt
\label{equation8}
\end{eqnarray}

The second canonical transformation is given by

\begin{eqnarray}
&\bar{\theta}_{1}&=\frac{(\theta_{1}+\theta_{0}+\theta_{-1})}{3} \,\,\,\,\, \bar{N}_{1}=N_{1}+N_{0}+N_{-1} \nonumber \\
&\bar{\theta}_{2}&=\theta_{0}-\frac{(\theta_{1}+\theta_{-1})}{2} \,\,\,\, \bar{N}_{2}=\frac{2}{3}N_{0}-\frac{1}{3}(N_{1}+N_{-1}) \\
&\bar{\theta}_{3}&=\theta_{1}-\theta_{-1} \,\,\,\,\,\,\,\,\,\,\,\,\,\,\,\,\,\,\,\,\,\, \bar{N}_{3}=\frac{(N_{1}-N_{-1})}{2} \nonumber
\label{equation9}
\end{eqnarray}

Two of these variables are, respectively, the mean number of atoms

\begin{eqnarray}
\bar{N}_{1}=\langle \vec{z}|\hat{N}|\vec{z}\rangle
\label{equation10}
\end{eqnarray}

\noindent and one-half the mean value of the component of the total hyperfine spin in the direction of the magnetic field

\begin{eqnarray}
\bar{N}_{3}=\frac{\langle\vec{z}|\hat{S}_{z}|\vec{z}\rangle}{2\hbar}
\label{equation11}
\end{eqnarray}

In terms of these new canonical variables the action becomes equal to

\begin{eqnarray}
S=\int_{t_{1}}^{t_{2}}\left( \sum_{\alpha}\hbar\bar{N}_{\alpha}\dot{\bar{\theta}}_{\alpha}-H_{0}(\bar{N}_{1},\bar{N}_{2},\bar{N}_{3},\bar{\theta}_{2})\right) dt
\label{equation12}
\end{eqnarray}

\noindent where the Hamiltonian is ciclic in the coordinates $\bar{\theta}_{1}$ and $\bar{\theta}_{3}$,

\begin{eqnarray}
H_{0}(\bar{N}_{1},\bar{N}_{2},\bar{N}_{3},\bar{\theta}_{2})= -2p\bar{N}_{3}+q\left( \frac{2}{3}\bar{N}_{1}-\bar{N}_{2}\right) \nonumber \\
+\frac{c_{0}}{2V}\bar{N}_{1}^{2}+\frac{c_{2}}{V}\left( 2\bar{N}_{3}^{2}+\frac{1}{3}(\bar{N}_{1}+\bar{N}_{2})\left( \frac{2}{3}\bar{N}_{1}-\bar{N}_{2}\right)\right. \nonumber \\
\left. +\sqrt{\left( \frac{2}{3}\bar{N}_{1}-\bar{N}_{2}\right)^{2} -4\bar{N}_{3}^{2}}\left(\frac{1}{3}\bar{N}_{1}+\bar{N}_{2} \right)\cos 2\bar{\theta}_{2}\right)
\label{equation13}
\end{eqnarray}

As stated before, the dynamics follows from the condition that the action (\ref{equation12}) is stationary, which leads to Hamilton equations of motion in these new canonical variables,

\begin{eqnarray}
\dot{\bar{\theta}}_{1}&=&\frac{\partial H_{0}}{\partial \bar{N}_{1}} \,\,\,\,\,\,\,\,\,\,\,\,\,\,\, \dot{\bar{N}}_{1}=0 \nonumber \\
\dot{\bar{\theta}}_{2}&=&\frac{\partial H_{0}}{\partial \bar{N}_{2}} \,\,\,\,\,\,\,\,\,\,\,\,\,\,\, \dot{\bar{N}}_{2}=-\frac{\partial H_{0}}{\partial \bar{\theta}_{2}} \\
\dot{\bar{\theta}}_{3}&=&\frac{\partial H_{0}}{\partial \bar{N}_{3}} \,\,\,\,\,\,\,\,\,\,\,\,\,\,\, \dot{\bar{N}}_{3}=0 \nonumber
\label{equation14}
\end{eqnarray}

The property that the hamiltonian is ciclic in $\bar{\theta}_{1}$ and $\bar{\theta}_{3}$ has two important consequences. One is that the mean-value of the total number of particles and of the component of the total spin of the condensate in the direction of the magnetic field are constants of the motion, $\bar{N}_{1}=N$, $2\hbar\bar{N}_{3}=\langle \vec{z}|S_{z}|\vec{z}\rangle$. The other is that the dynamics involves only one degree of freedom

\begin{eqnarray}
\dot{\bar{\theta}}_{2}=\frac{\partial H_{0}}{\partial \bar{N}_{2}} \,\,\,\,\,\,\,\,\,\,\,\,\,\,\, \dot{\bar{N}}_{2}=-\frac{\partial H_{0}}{\partial \bar{\theta}_{2}}
\label{equation15}
\end{eqnarray}

To simplify the equations of motion (\ref{equation15}) we define new variables which are the old variables divided by the number of particles, $\bar{n}_{\alpha}=\frac{\bar{N}_{\alpha}}{N}$, to write the equations of motion as

{\small
\begin{eqnarray}
&\hbar& \dot{\bar{n}}_{2}=2c_{2}\rho\sqrt{\left(\frac{2}{3}-\bar{n}_{2} \right)^{2}-4\bar{n}_{3}^{2}}\left(\frac{1}{3}+\bar{n}_{2} \right)\sin 2\bar{\theta}_{2} \nonumber \\
-&\hbar&\dot{\bar{\theta}}_{2}=q+c_{2}\rho\left(\left(2\bar{n}_{2}-\frac{1}{3} \right)\right.  \nonumber  \\
&+&\left.  \frac{\left(\frac{2}{3}-\bar{n}_{2} \right)\left(2\bar{n}_{2}-\frac{1}{3} \right)+4\bar{n}_{3}^{2}}{\sqrt{\left(\frac{2}{3}-\bar{n}_{2} \right)^{2}-4\bar{n}_{3}^{2}}}\cos 2\bar{\theta}_{2} \right)
\label{equation16}
\end{eqnarray}
}

\noindent where $\rho=\frac{N}{V}$ is the density of the condensate.

The equations (\ref{equation16}) are the analog of the mean-field classical equations of motion that describes atoms in two states coupled by a Josephson-type term \cite{Smerzi}.

From equations (\ref{equation16}), the following general properties of the equations of motion emerges:

1) The population dynamics is independent of the strength of the linear term of the Zeeman energy and of the spin-independent component of the atom-atom interaction. Indeed, only the phase $\bar{\theta}_{3}$ depends on $p$ and only the phase $\bar{\theta}_{1}$ depends on $c_{0}$.

2) By a proper choice of time scale, $\tau=\frac{\hbar}{|c_{2}\rho|}$, the population dynamics, in the limit $q=0$, is independent of the magnitude of $c_{2}$, depending only on its sign.
In general, that is when $q\neq 0$, the population dynamics depends on the ratio $\frac{q}{c_{2}\rho}$ and on the sign of $c_{2}$.

\section{Properties of the equations of motion}
\subsection{Equilibrium configurations}

From the equations of motion, equations (\ref{equation16}), we see that the equations which determine the equilibrium configurations are

{\small
\begin{subequations}
\begin{eqnarray}
\sqrt{\left(\frac{2}{3}-\bar{n}_{2} \right)^{2}-4\bar{n}_{3}^{2}}\left(\frac{1}{3}+\bar{n}_{2} \right)\sin 2\bar{\theta}_{2}=0
\label{equation17a}
\end{eqnarray}
\begin{eqnarray}
\frac{q}{c_{2}\rho}+\left(2\bar{n}_{2}-\frac{1}{3} \right)+ \,\,\,\,\,\,\,\,\,\,\,\,\,\,\,\,\,\,\,\,\,\,\,\,\,\,\,\,\,\,\,\,\,\,\,\,\,\,\,\,\,\,\,\,\,\, \nonumber \\
+\frac{\left(\frac{2}{3}-\bar{n}_{2} \right)\left(2\bar{n}_{2}-\frac{1}{3} \right) +4\bar{n}_{3}^{2}}{\sqrt{\left(\frac{2}{3}-\bar{n}_{2} \right)^{2}-4\bar{n}_{3}^{2}}}\cos 2\bar{\theta}_{2}=0
\label{equation17b}
\end{eqnarray}
\end{subequations}
}

In equations (\ref{equation16}) and (\ref{equation17a},\ref{equation17b}) the constant of the motion $\bar{n}_{3}$ is defined in the interval $-\frac{1}{2}<\bar{n}_{3}<\frac{1}{2}$ and the dynamic variable $\bar{n}_{2}$ in the interval $-\frac{1}{3}<\bar{n}_{2}<\frac{2}{3}-2|\bar{n}_{3}|$. In our discussion of the solutions of the equilibrium equations we consider separately the cases $\bar{n}_{3}\neq 0$ and $\bar{n}_{3}=0$.

\vspace{0.5cm}

{\bf a)} $\bar{n}_{3}\neq 0$

\vspace{0.25cm}

We have solutions which depends on the phase $\bar{\theta}_{2}$ and solutions which are independent of $\bar{\theta}_{2}$

\vspace{0.5cm}

{\bf a1)} Solution which depends on $\bar{\theta}_{2}$ with $\cos 2\bar{\theta}_{2}=1$.

\vspace{0.25cm}

In this case the equilibrium value of $\bar{n}_{2}$ is given by equation (\ref{equation17b}) with $\cos 2\bar{\theta}_{2}=1$. This equation have one solution in the interval $-\infty<\frac{q}{c_{2}\rho}<1+\sqrt{1-(2\bar{n}_{3})^{2}}$. When $\frac{q}{c_{2}\rho}\rightarrow -\infty$ the equilibrium value of $\bar{n}_{2}$ approaches the upper boundary, $\bar{n}_{2}=\frac{2}{3}-2|\bar{n}_{3}|$, for which the fraction of atoms occupying the hyperfine levels are $n_{1}=|\bar{n}_{3}|+\bar{n}_{3}$, $n_{0}=1-2|\bar{n}_{3}|$, $n_{-1}=|\bar{n}_{3}|-\bar{n}_{3}$. On the other hand when $\frac{q}{c_{2}\rho}=1+\sqrt{1-(2\bar{n}_{3})^{2}}$, $\bar{n}_{2}$ is at the lower boundary, $\bar{n}_{2}=-\frac{1}{3}$, in which case the fraction of atoms occupying the hyperfine levels are $n_{1}=\frac{1}{2}(1+2\bar{n}_{3})$, $n_{0}=0$, $n_{-1}=\frac{1}{2}(1-2\bar{n}_{3})$. 

When we neglect the quadratic term of the Zeeman energy, that is $q=0$, the equilibrium value of 
$\bar{n}_{2}$ is, $\bar{n}_{2}=\frac{1}{2}\left(\frac{1}{3}-(2\bar{n}_{3})^{2} \right)$ and the occupation fractions are $n_{1}=\frac{1}{4}(1+2\bar{n}_{3})^{2}$, $n_{0}=\frac{1-(2\bar{n}_{3})^{2}}{2}$, $n_{-1}=\frac{1}{4}(1-2\bar{n}_{3})^{2}$.

\vspace{ 0.5cm}

{\bf a2)} Solution which depends on $\bar{\theta}_{2}$ with $\cos 2\bar{\theta}_{2}=-1$

\vspace{0.25cm}

The equilibrium value of $\bar{n}_{2}$ is now given by equation (\ref{equation17b}) with $\cos 2\bar{\theta}_{2}=-1$. This equation has one solution in the interval $\frac{q}{c_{2}\rho}>1-\sqrt{1-(2\bar{n}_{3})^{2}}$. When $\frac{q}{c_{2}\rho}=1-\sqrt{1-(2\bar{n}_{3})^{2}}$ $\bar{n}_{2}$ is at the lower boundary $\bar{n}_{2}=-\frac{1}{3}$. On the other hand when $\frac{q}{c_{2}\rho}\rightarrow \infty$ it approaches the upper boundary $\bar{n}_{2}=\frac{2}{3}-2|\bar{n}_{3}|$.

\vspace{0.5cm}

{\bf a3)} Solution which does not depend on the phase $\bar{\theta}_{2}$

\vspace{0.25cm}

In this case $\bar{n}_{2}$ is at the lower boundary $\bar{n}_{2}=-\frac{1}{3}$, independently of the value of $\frac{q}{c_{2}\rho}$. 

\begin{figure}
\psfrag{q1}{$\frac{q}{c_{2}\rho}$}
\psfrag{q}{}
\psfrag{n2}{$\bar{n}_{2}$}
\psfrag{regiao1}{}
\psfrag{regiao2}{}
\psfrag{regiao3}{}
\includegraphics[width=9cm]{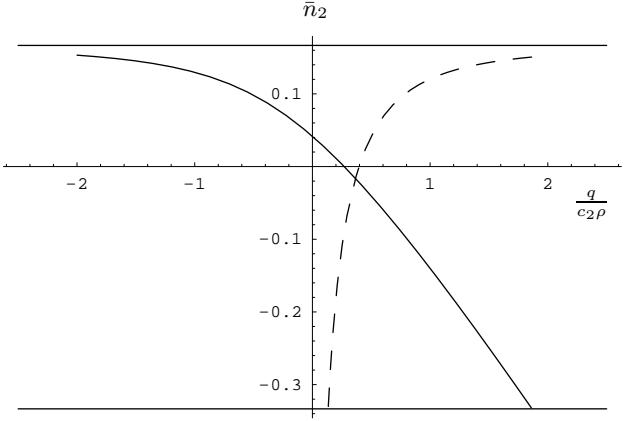}
\caption{\label{figure1} Equilibrium values of $\bar{n}_{2}$ as function of $\frac{q}{c_{2}\rho}$ for $\bar{n}_{3}=\frac{1}{4}$. Solid curve is {\bf a1)}, dashed is {\bf a2)}, and the lower boundary $\bar{n}_{2}=-\frac{1}{3}$ is {\bf a3)}. The horizontal straight lines are the boundaries of $\bar{n}_{2}$.}
\end{figure}

These properties are illustrated in FIG.\ref{figure1} where we plot the equilibrium values of $\bar{n}_{2}$ as a function of $\frac{q}{c_{2}\rho}$ for $\bar{n}_{3}=\frac{1}{4}$.

\begin{table*}
\caption{\label{table1}Equilibrium configurations for different parameter domains. $\bar{n}_{3}\neq 0$, antiferromagnetic case, $c_{2}\rho>0$.}
\begin{ruledtabular}
\begin{tabular}{cc}

 $\frac{q}{|c_{2}\rho|}<1-\sqrt{1-(2\bar{n}_{3})^{2}}$&{\bf a1)} maximum; {\bf a3)} minimum \\
 .&. \\
 $1-\sqrt{1-(2\bar{n}_{3})^{2}}<\frac{q}{|c_{2}\rho|}<1+\sqrt{1-(2\bar{n}_{3})^{2}}$&{\bf a1)} maximum; {\bf a2)} minimum; {\bf a3)} undefined\footnote{The classification undefined means that when we leave the line defining the corresponding boundary the energy increases or decreases depending of the value of the phase $\bar{\theta}_{2}$.}\\
 $.$&$.$\\
 $\frac{q}{|c_{2}\rho|}>1+\sqrt{1-(2\bar{n}_{3})^{2}}$& {\bf a2)} minimum; {\bf a3)} maximum
\end{tabular}
\end{ruledtabular}
\end{table*}

\begin{table*}
\caption{\label{table2}Equilibrium configurations for different parameter domains. $\bar{n}_{3}\neq 0$, ferromagnetic case, $c_{2}\rho<0$.}
\begin{ruledtabular}
\begin{tabular}{cc}

 $\frac{q}{|c_{2}\rho|}>-\left(1-\sqrt{1-(2\bar{n}_{3})^{2}}\right)$&{\bf a1)} minimum; {\bf a3)} maximum \\
 .&. \\
 $-\left(1+\sqrt{1-(2\bar{n}_{3})^{2}}\right)<\frac{q}{|c_{2}\rho|}<-\left(1-\sqrt{1-(2\bar{n}_{3})^{2}}\right)$&{\bf a1)} minimum; {\bf a2)} maximum; {\bf a3)} undefined\\
 $.$&$.$\\
 $\frac{q}{|c_{2}\rho|}<-\left(1+\sqrt{1-(2\bar{n}_{3})^{2}}\right)$& {\bf a2)} maximum; {\bf a3)} minimum
\end{tabular}
\end{ruledtabular}
\end{table*}

\begin{table*}
\caption{\label{table3}Equilibrium configurations for different parameter domains. $\bar{n}_{3}=0$, antiferromagnetic limit, $c_{2}\rho>0$.}
\begin{ruledtabular}
\begin{tabular}{cc}

 $\frac{q}{|c_{2}\rho|}<-2$&{\bf b3)} minimum; {\bf b4)} maximum \\
 .&. \\
 $-2<\frac{q}{|c_{2}\rho|}<0$&{\bf b1)} maximum; {\bf b3)} minimum; {\bf b4)} undefined \\
 $.$&$.$\\
 $q=0$& {\bf b1)} maximum; {\bf b2)},{\bf b3)},{\bf b4)} degenerate minimum\\
 .&.\\
 $0<\frac{q}{|c_{2}\rho|}<2$& {\bf b1)} maximum; {\bf b3)} undefined; {\bf b4)} minimum\\
 .&.\\
 $\frac{q}{|c_{2}\rho|}>2$&{\bf b3)} maximum; {\bf b4)} minimum \\
\end{tabular}
\end{ruledtabular}
\end{table*}

\begin{table*}
\caption{\label{table4}Equilibrium configurations for different parameter domains. $\bar{n}_{3}=0$, ferromagnetic limit, $c_{2}\rho<0$.}
\begin{ruledtabular}
\begin{tabular}{cc}

 $\frac{q}{|c_{2}\rho|}>2$&{\bf b3)} maximum; {\bf b4)} minimum \\
 .&. \\
 $0<\frac{q}{|c_{2}\rho|}<2$&{\bf b1)} minimum; {\bf b3)} maximum; {\bf b4)} undefined \\
 $.$&$.$\\
 $q=0$& {\bf b1)} minimum; {\bf b2)},{\bf b3)},{\bf b4)} degenerate maximum\\
 .&.\\
$-2<\frac{q}{|c_{2}\rho|}<0$ & {\bf b1)} minimum; {\bf b3)} undefined;{\bf b4)} maximum\\
.&. \\
$\frac{q}{|c_{2}\rho|}<-2$&{\bf b3)} minimum; {\bf b4)} maximum
\end{tabular}
\end{ruledtabular}
\end{table*}

\vspace{0.5cm}

{\bf b)} $\bar{n}_{3}=0$

\vspace{0.25cm}

In this case the equilibrium equations, (\ref{equation17a},\ref{equation17b}), reduces to

\begin{eqnarray}
\left(\frac{2}{3}-\bar{n}_{2} \right)\left(\frac{1}{3}+\bar{n}_{2} \right)\sin 2\bar{\theta}_{2}=0 \nonumber \\
\frac{q}{c_{2}\rho}+\left(2\bar{n}_{2}-\frac{1}{3} \right)\left(1+\cos 2\bar{\theta}_{2} \right)=0 \nonumber
\end{eqnarray}

Again, we have two classes of solutions, dependent and independent of the phase $\bar{\theta}_{2}$.

{\bf b1)} Solution which depends on the phase $\bar{\theta}_{2}$ with $\cos 2\bar{\theta}_{2}=1$

\vspace{0.25cm}

The equilibrium value of $\bar{n}_{2}$ is $\bar{n}_{2}=\frac{1}{2}\left(\frac{1}{3}-\frac{q}{2c_{2}\rho} \right)$ in the interval $-2<\frac{q}{c_{2}\rho}<2$. In this case the occupation fraction of the hyperfine levels are $n_{1}=n_{-1}=\frac{1}{4}\left(1+\frac{q}{2c_{2}\rho} \right)$, $n_{0}=\frac{1}{2}\left(1-\frac{q}{2c_{2}\rho} \right)$.

\vspace{0.5cm}

{\bf b2)} Solution which depends on the phase $\bar{\theta}_{2}$ with $\cos 2\bar{\theta}_{2}=-1$

\vspace{0.25cm}

In this case there is a solution only at $q=0$ and it is the line $\cos 2\bar{\theta}_{2}=-1$, $-\frac{1}{3}<\bar{n}_{2}<\frac{2}{3}$.

When $\bar{n}_{3}=0$, there are two solutions which does not depend on the phases.

\vspace{0.5cm}

{\bf b3)} One is the lower boundary $\bar{n}_{2}=-\frac{1}{3}$ in which case the occupation fractions are $n_{1}=n_{-1}=\frac{1}{2}$, $n_{0}=0$.

\vspace{0.5cm}

{\bf b4)} The other is the upper boundary $\bar{n}_{2}=\frac{2}{3}$ in which case the occupation fractions are $n_{1}=n_{-1}=0$, $n_{0}=1$. The solutions {\bf b3)} and {\bf b4)} exist for any value of $\frac{q}{c_{2}\rho}$.

\begin{figure}
\psfrag{q1}{$\frac{q}{c_{2}\rho}$}
\psfrag{q}{}
\psfrag{n2}{$\bar{n}_{2}$}
\psfrag{regiao1}{}
\psfrag{regiao2}{}
\psfrag{regiao3}{}
\includegraphics[width=9cm]{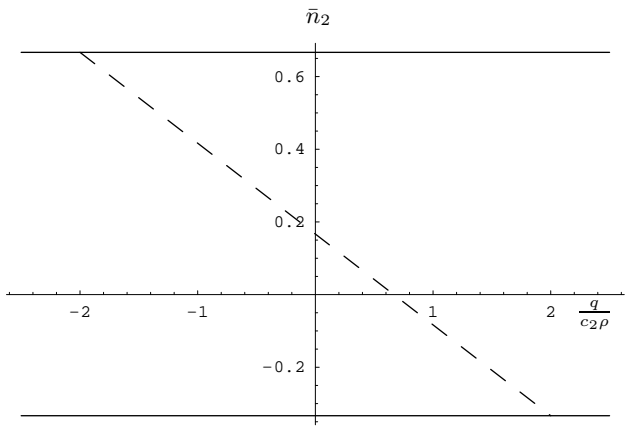}
\caption{\label{figure2} Equilibrium values of $\bar{n}_{2}$ as function of $\frac{q}{c_{2}\rho}$ for $\bar{n}_{3}=0$. The dashed straight line is {\bf b1)}. The vertical straight line $q=0$ is {\bf b2)}. The horizontal straight lines are the boundaries of $\bar{n}_{2}$.}
\end{figure}

The properties of the $\bar{n}_{3}=0$ equilibrium configurations are illustrated in FIG.\ref{figure2} , where we plot the equilibrium values of $\bar{n}_{2}$ as a function of $\frac{q}{c_{2}\rho}$.

\begin{figure}
\psfrag{energia}{Energy}
\psfrag{n2}{$\bar{n}_{2}$}
\psfrag{theta}{$\bar{\theta}_{2}$}
\includegraphics[width=7cm]{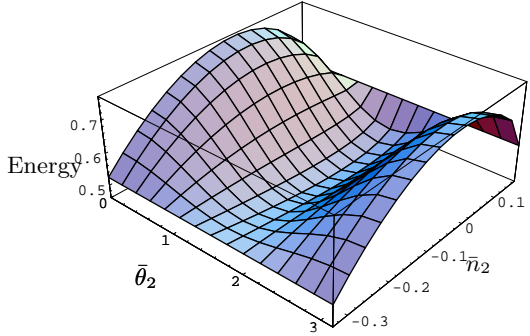}
\includegraphics[width=7cm]{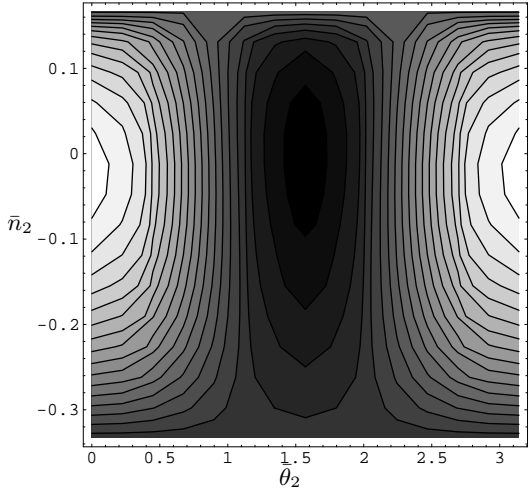}
\caption{\label{figure3} Top: energy surface as function of $\bar{\theta}_{2}$ and $\bar{n}_{2}$. Bottom: contour plot of the energy surface in the $\bar{\theta}_{2} \times \bar{n}_{2}$ plane. In these plots $\bar{n}_{3}=\frac{1}{4}$ and $\frac{q}{|c_{2}\rho|}=0.4$, in the antiferromagnetic limit. The energy is in units of  $|c_{2}\rho|$. Darker colors means lower energy. }
\end{figure}

\begin{figure}
\psfrag{energia}{Energy}
\psfrag{n2}{$\bar{n}_{2}$}
\psfrag{theta}{$\bar{\theta}_{2}$}
\includegraphics[width=7cm]{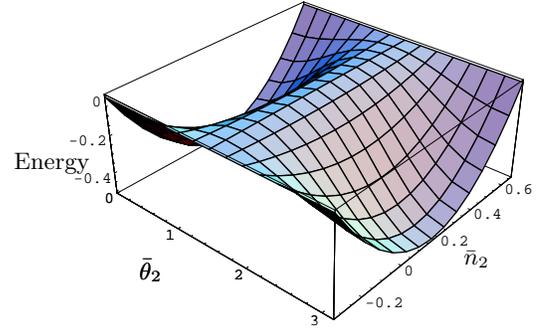}
\includegraphics[width=7cm]{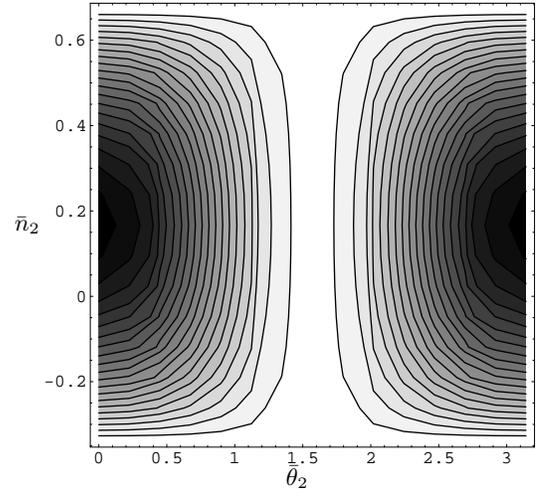}
\caption{\label{figure4} Top: energy surface as function of $\bar{\theta}_{2}$ and $\bar{n}_{2}$. Bottom: contour plot of the energy surface in the $\bar{\theta}_{2} \times \bar{n}_{2}$ plane. In this plots $\bar{n}_{3}=0$ and $\frac{q}{|c_{2}\rho|}=0$, in the ferromagnetic limit. The energy is in units of $|c_{2}\rho|$.}
\end{figure}

We would like to point out that it can be shown that the phase dependent solutions of the equilibrium equations are the roots of the third order equation,

\begin{eqnarray}
&4&\frac{q}{c_{2}\rho}f_{0}^{3}-\frac{q}{c_{2}\rho}\left(2+\frac{q}{c_{2}\rho}\right)f_{0}^{2}-(2\bar{n}_{3})^{2}2\left(1+2\frac{q}{c_{2}\rho} \right)f_{0} \nonumber \\
&+&(2\bar{n}_{3})^{2}\left(\left(1+\frac{q}{c_{2}\rho}\right)^{2}+(2\bar{n}_{3})^{2} \right) =0 \,\,\,\,\, f_{0}\neq 2|\bar{n}_{3}|
\label{equation18}
\end{eqnarray}

\noindent with $\bar{n}_{2}=\frac{2}{3}-f_{0}$, where $f_{0}$ is the fraction of atoms outside the $m_{f}=0$ hyperfine level, $f_{0}=n_{1}+n_{-1}$.

We can find analytical expressions for the roots of equation (\ref{equation18}) but they are not particularly illuminating and we will not write them here.

A summary of our discussion is displayed in the tables (\ref{table1},\ref{table2},\ref{table3},\ref{table4}), where we consider separately the antiferromagnetic and the ferromagnetic limits. From the tables we can easily determine what are the ground state configurations for the different parameter domains and in the two limits considered above. 

Ground state configurations have been investigated in reference \cite{Ho}, in the absence of the magnetic field $(p=q=0)$. They observe that the (degenerate) ground state in the antiferromagnetic limit are the polar states and in the ferromagnetic limit, the ferromagnetic states. 

The polar states of \cite{Ho} have $\bar{n}_{3}=0$ and are the states {\bf b2)},{\bf b3)}, and {\bf b4)} at $q=0$ (see Table III). When $\bar{n}_{3}=0$, the ground state in the ferromagnetic limit is the state {\bf b1)} which is equal to the ferromagnetic state of \cite{Ho} with $\bar{n}_{3}=0$.

When $\bar{n}_{3}\neq 0$, the states {\bf a1)} at $q=0$ are equal to the ferromagnetic states of \cite{Ho} and are the ground states in the ferromagnetic limit (see Table II). On the other hand, the ground state in the antiferromagnetic limit is the state {\bf a3)}, which is not a polar state.

Reference \cite{Ohmi} investigate the ground state configurations  in the presence of the magnetic field and neglecting the quadratic term of the Zeeman energy. As shown in this paper, in this case the ground state configurations coincide with the ground state configurations in the absence of the magnetic field, a fact overlooked in \cite{Ohmi}. Besides a state with maximum value of $\bar{n}_{3}$, $\bar{n}_{3}=\frac{1}{2}$, \cite{Ohmi} identify only the spinor {\bf a3)}, which is the ground state just in the antiferromagnetic limit (see Table I).

The general case when we consider both the linear and quadratic terms of the Zeeman energy have been studied in reference \cite{Stenger}. Our approach differs from \cite{Stenger}, in the sense that we take explicitly into account the constraints of axial symmetry which simplifies considerably the discussion.


\subsection{Miscibility}

One question that we can adress is the miscibility of the hyperfine
 components in the ground state spinors \cite{Stenger}.

Consider first the case $\bar{n}_{3}\neq 0$ and the antiferromagnetic
limit. For $q<1-\sqrt{1-(2\bar{n}_{3})^{2}}$, the ground state spinor
is {\bf a3)}, where only the $m_{f}=\pm 1$  components are miscible ,
the population of the $m_{f}=0$ component being null. For
$\frac{q}{|c_{2}\rho|}>1-\sqrt{1-(2\bar{n}_{3})^{2}}$, the ground state
spinor is {\bf a2)}, in which case the three components are generally
miscible. Actually the miscibility of the hyperfine components depends
on the ratio $\frac{q}{|c_{2}\rho|}$. When $\frac{q}{|c_{2}\rho|}$
approaches its lowest value, $\bar{n}_{2}$ is near the lower boundary, 
$\bar{n}_{2}=-\frac{1}{3}$, therefore  the $m_{f}=0$ component is
practically  immiscible with the $m_{f}=\pm 1$ components. When 
$\frac{q}{|c_{2}\rho|}$ increases, {\bf a2)} approaches the upper boundary
$\bar{n}_{2}=\frac{2}{3}-2|\bar{n}_{3}|$, in which case  
 the $m_{f}=0$ component mixes predominantly with the $m_{f}=+1(-1)$ component depending on the sign of $\bar{n}_{3}$, $\bar{n}_{3}>0 (<0)$, the population of the third component being  negligible.

The next case is $\bar{n}_{3}\neq 0 $ and the  ferromagnetic limit. For
 $\frac{q}{|c_{2}\rho|}<-\left(1+\sqrt{1-(2\bar{n}_{3})^{2}} \right)$,
 the ground state  spinor is {\bf a3)}, where only the $m_{f}=\pm 1$
 components are miscible, the population of the $m_{f}=0$ component
 being null. On the other hand, when
 $\frac{q}{|c_{2}\rho|}>-\left(1+\sqrt{1-(2\bar{n}_{3})^{2}} \right)$
 the ground state spinor is {\bf a1)}. As in the antiferromagnetic
 limit, in this case 
 the three hyperfine components are generally miscible, the degree of
 miscibility depending on  the ratio $\frac{q}{|c_{2}\rho|}$. When this
 ratio approches its lowest value, $\bar{n}_{2}$ is near the lower
 boundary, $\bar{n}_{2}=-\frac{1}{3}$, and  the $m_{f}=0$ component is
 practically immiscible with the $m_{f}=\pm 1$ component. On the other
 hand, when $\frac{q}{|c_{2}\rho|}$ increases approachig $+ \infty $, $\bar{n}_{2} $ is near the upper boundary, $\bar{n}_{2}=
 \frac{2}{3}- 2\bar{n}_{3}$, in which case the $m_{f}=0$ component mixes
 predominantly with the $m_{f}=+1(-1)$ component depending on the sign
 of $\bar{n}_{3}$, $\bar{n}_{3}>0 (<0)$, the population of the
 third component being  negligible.
 
 The only case left is when $\bar{n}_{3}=0$. Consider first the antiferromagnetic limit. For $\frac{q}{|c_{2}\rho|}<0$ the ground state spinor is {\bf b3)}, for which $\bar{n}_{2}$ is at the lower boundary, $\bar{n}_{2}=-\frac{1}{3}$. In this spinor we have equal population of the $m_{f}=\pm1$ components, the population of the $m_{f}=0$ component being null. On the other hand, for $\frac{q}{|c_{2}\rho|}>0$, the ground state spinor is {\bf b4)} for which $\bar{n}_{2}$ is at the upper boundary, $\bar{n}_{2}=\frac{2}{3}$. In this spinor all atoms are in the $m_{f}=0$ state. 
 
 The ferromagnetic limit is richer than the previous one. For $\frac{q}{|c_{2}\rho|}<-2$ the ground state spinor is {\bf b3)}. However, for $-2<\frac{q}{|c_{2}\rho|}<2$, the ground state spinor is {\bf b1)}, where the three components are generally miscible. Actually, it changes from a spinor where practically only the $m_{f}=\pm 1$ are miscible, the population of the $m_{f}=0$ component being negligible, near the lowest value of $\frac{q}{|c_{2}\rho|}$ to one where practically only the component $m_{f}=0$ is populated near the highest value of $\frac{q}{|c_{2}\rho}|$. For $\frac{q}{|c_{2}\rho|}>2$ the ground state spinor is {\bf b4)}.

\section{Dynamics}
 
\begin{figure}
\psfrag{n2}{$\bar{n}_{2}$}
\psfrag{t}{$t$}
\psfrag{theta2}{$\bar{\theta}_{2}$}
\subfigure[]{
\begin{minipage}[t]{.48\linewidth}
\includegraphics[width=\linewidth]{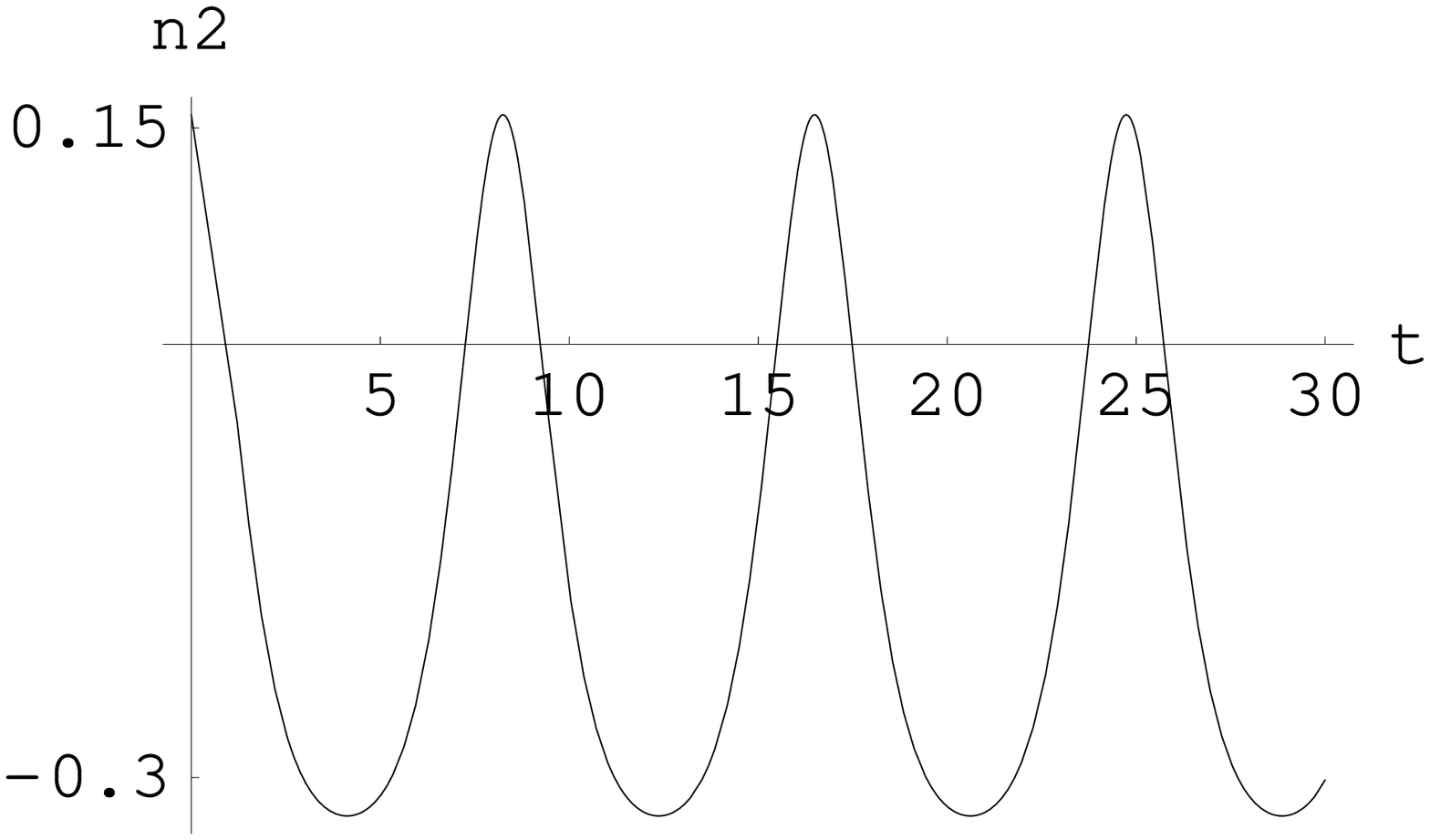}
\end{minipage}\hfill}
\subfigure[]{\begin{minipage}[t]{.48\linewidth}
\includegraphics[width=\linewidth]{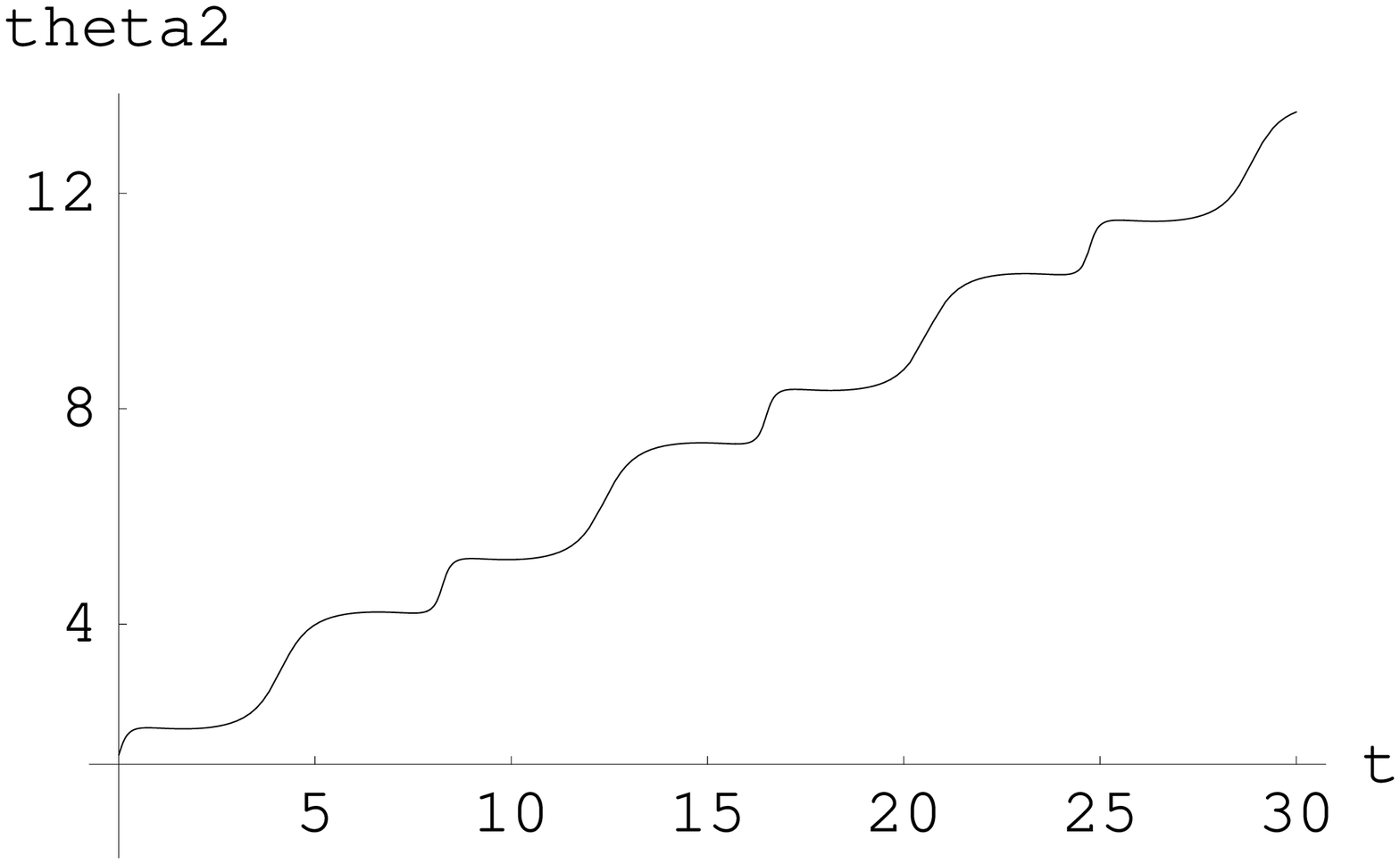}
\end{minipage}\hfill}
\subfigure[]{\begin{minipage}[b]{.48\linewidth}
\includegraphics[width=\linewidth]{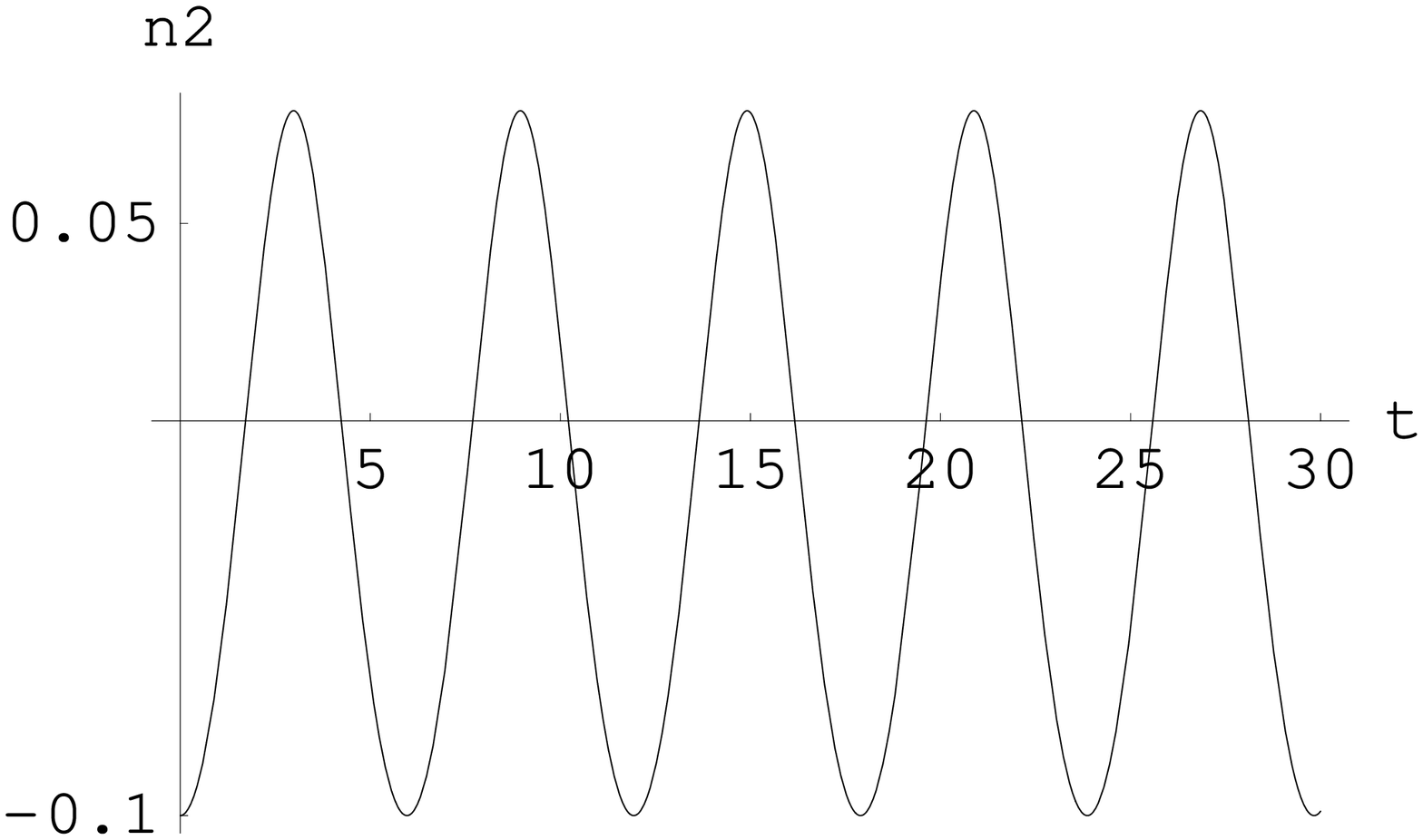}
\end{minipage}\hfill}
\subfigure[]{\begin{minipage}[b]{.48\linewidth}
\includegraphics[width=\linewidth]{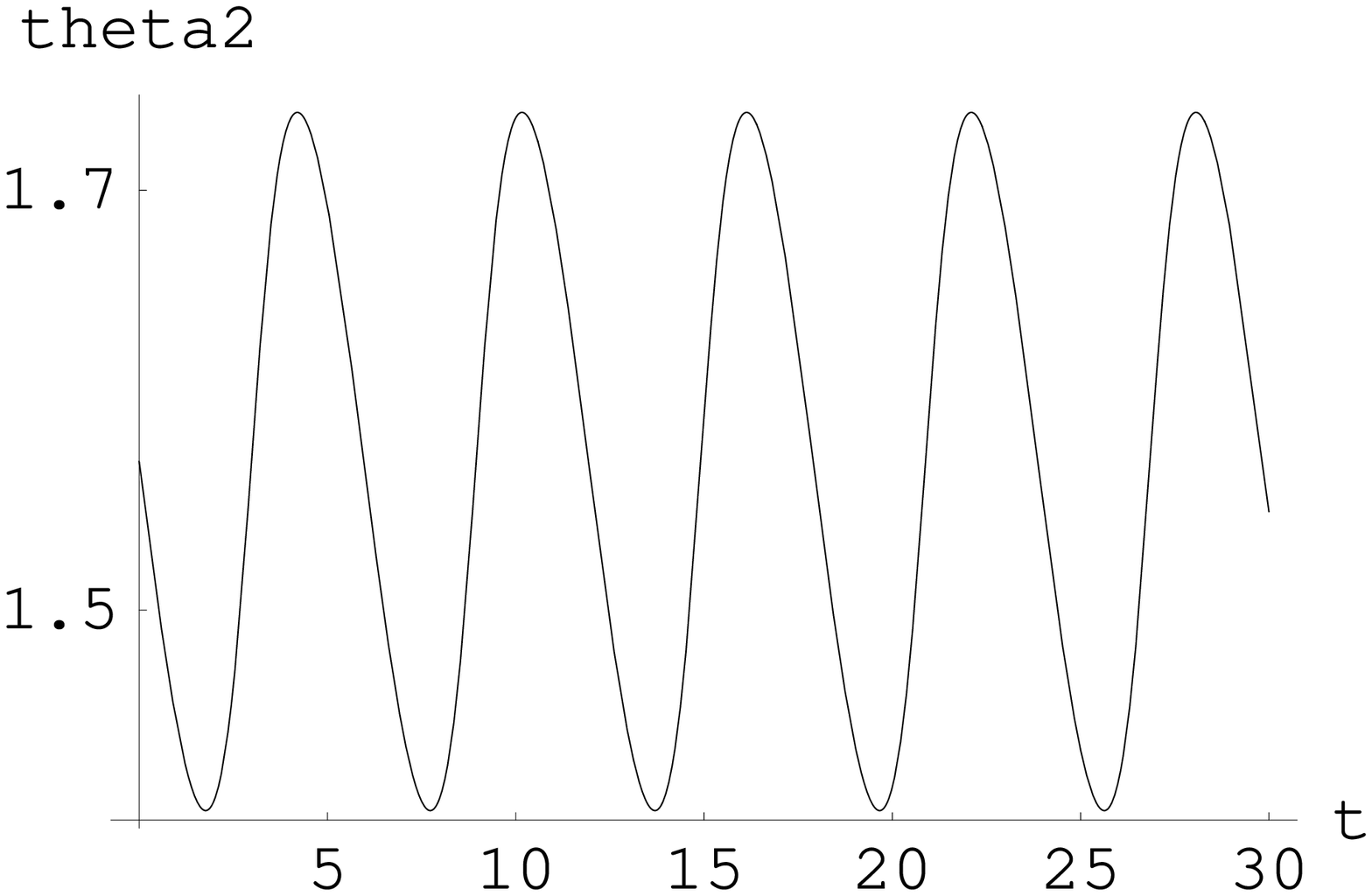}
\end{minipage}\hfill}
\caption{{\bf (a)}$\bar{n}_{2}$ ,{\bf (b)}$\bar{\theta}_{2}$ as function of time, in the case of rotation; {\bf (c)}$\bar{n}_{}2$, {\bf (d)}$\bar{\theta}_{2}$ as funtion of time, in the case of libration. Notice that $\bar{n}_{2}$ is always periodic. The time is in units of $\frac{\hbar}{|c_{2}\rho|}$ and the angles, in radians. The parameters are equal to the ones in FIG.\ref{figure3}. The initial conditions are: $\bar{n}_{2}=0.159$ and $\bar{\theta}_{2}=\frac{\pi}{2}$ for rotation, $\bar{n}_{2}=-0.1$ and $\bar{\theta}_{2}=\frac{\pi}{2}$ for libration.}
\label{figure5}
\end{figure}

The qualitative features of the population dynamics can be easily visualized if we make portraits of the contour curves $\frac{H_{0}}{N}=$constant in the phase space plane $\bar{\theta}_{2}\times \bar{n}_{2}$. Examples are shown in FIG.\ref{figure3} and 
FIG.\ref{figure4}. We see that $\bar{n}_{2}(t)$ is always a periodic function of time. The motion can be a libration when $\bar{\theta}_{2}$ is a limited function of time and a rotation in which case $\bar{\theta}_{2}$ always increases (decreases) as a function of time. Examples of these behaviors are displayed in FIG.\ref{figure5}.

\noindent

\section{Summary and conclusions}

In summary we have studied in this paper the classical dynamics that underlies the mean-field description of an homogeneous mixture of spinor $F=1$ condensate in an external magnetic field. 

As a consequence of number conservation and axial symmetry in spin space this dynamics is integrable. The equations of motion show that the population dynamics depends only on the quadratic term of the Zeeman energy and on the strength of the spin-dependent component of the atom-atom interaction.

For a fixed mean-value of the component of the condensate spin in the direction of the magnetic field we determine the equilibrium configurations as a function of the ratio $\frac{q}{c_{2}\rho}$. We also make a detailed discussion of the miscibility of the three hyperfine components in the ground state spinor as a function of $\frac{q}{c_{2}\rho}$. Our studies reveal the absence of metastability in the sense that there is no two local minima configurations in the parameter domain. We have shown that outside the equilibrium, the populations are always a periodic function of time, where the periodic motion can be a libration or a rotation. In the first case the phase $\bar{\theta}_{2}$ is always limited whereas in the second case it always increase (decrease), as function of time.  

Finally we would like to remark that the restriction to a homogeneous mixture is not only of academic interest \cite{Ho,Ohmi}. Besides being a guide to what happens in the case of trapped condensates, classical Hamiltonians of the type considered in this paper emerges in a mean-field description of the quantum single-mode approximation for spinor condensates in a trap \cite{Law}. 

DRR would like to acknowledge financial support from CNPq, and EJVP partial support from CNPq.

\end{document}